# With foundation-model features, aggregator complexity does not improve cross-hospital discrimination in gastric whole-slide image classification

Nheng Vanchhay[1], Ali Hussain[1], Khan Ziaullah[1], Moon-Il Joo[1], Md Ariful Islam Mozumder[1], Rashedul Islam Sumon[1], Won-Ae Lee[2], Min-Sun Cho[3], Hyeong-Chan Shin[4], Yujun Park[5], Baek Hwan Cho[6], Nam-Hoon Cho[7], Heejung An[5], Heung-Kook Choi[8], Haeyoun Kang[5], Hyunki Kim[9], Hee Cheol Kim[1*]

[1]College of AI Convergence, Institute of Digital Anti-Aging Healthcare, u-AHRC, Inje University, 197 Inje-ro, Gimhae, 50834, Gyeongsangnam-do, Republic of Korea.

[2]Department of Pathology, Dankook University Hospital, 201 Manghyang-ro, Dongnam-gu, Cheonan, 31116, Chungcheongnam-do, Republic of Korea.

[3]Department of Pathology, Ewha Womans University Seoul Hospital, 260 Gonghang-daero, Gangseo-gu, Seoul, 07804, Republic of Korea.

[4]Department of Pathology, Keimyung University Dongsan Medical Center, 1035 Dalgubeol-daero, Dalseo-gu, Daegu, 42601, Republic of Korea.

[5]Department of Pathology, CHA Bundang Medical Center, CHA University, 59 Yatap-ro, Bundang-gu, Seongnam-si, 13496, Gyeonggi-do, Republic of Korea.

[6]Department of Biomedical Informatics, CHA University School of Medicine, CHA University, 335 Pangyo-ro, Bundang-gu, Seongnam-si, 13488, Gyeonggi-do, Republic of Korea.

[7]MTS Company Co., Ltd., 8F, 434 Samseong-ro, Gangnam-gu, Seoul, 06178, Republic of Korea.

[8]Department of Computer Engineering, Inje University, 197 Inje-ro, Gimhae, 50834, Gyeongsangnam-do, Republic of Korea.

[9]Department of Pathology, Yonsei University College of Medicine, 50-1 Yonsei-ro, Seodaemun-gu, Seoul, 03722, Republic of Korea.

[*]Corresponding author. E-mail: heeki@inje.ac.kr

Contributing authors: nhengvanchhay@live.inje.ac.kr; alihussain4541@live.inje.ac.kr; Ziakhan@live.inje.ac.kr; joomi@inje.ac.kr; arifulimdream@live.inje.ac.kr; walee@dankook.ac.kr; mcho1124@ewha.ac.kr; chan@dsmc.or.kr; yjpark7202@chamc.co.kr; baekhwan.cho@cha.ac.kr; cho1988@mtsco.co.kr; hjahn@cha.ac.kr; cschk@inje.ac.kr; hykang@cha.ac.kr; KIMHYUNKI@yuhs.ac.

## Abstract

Whole-slide image classifiers are usually developed within one hospital and often go untested at unseen hospitals. We assembled 5,036 haematoxylin-and-eosin-stained gastric slides from five Korean hospitals, scanned at 20× and 40×, and compared multiple-instance learning aggregators under leave-one-hospital-out validation on frozen pathology foundation-model features. A hierarchical patch–region–slide graph aggregator (CHIME-MIL) with site alignment and two-scale fusion reached a mean area under the receiver operating characteristic curve (AUC) of 0.837, above six conventional attention and transformer aggregators trained without these steps. Under a matched recipe it did not exceed attention-based pooling, and global max pooling given the same alignment and fusion was better (AUC 0.855, $P = 0.0002$). The alignment, which subtracts each hospital's mean feature vector without labels, onboarded a new hospital from ten unlabelled slides and cost 0.002 AUC without held-out-hospital data, but its accuracy gains were small and inconsistent. The model had low top-label calibration error (0.053) but under-confident malignancy probabilities, and removing its most-attended patches lowered the malignancy probability far more than removing random patches. With these features, more complex aggregators did not improve discrimination; what a deployable system adds is label-free onboarding, calibration checks and auditable evidence.

## Introduction

Histopathological examination of haematoxylin and eosin (H&E)-stained whole slide images (WSIs) is the diagnostic gold standard in oncology, but manual review is time-consuming, subject to inter-observer variability, and strained by rising case volumes. Computational pathology can automate slide-level classification at clinical scale, with deep models matching pathologist-level performance under controlled conditions [1]. Because WSIs exceed $10^9$ pixels, the standard pipeline tiles each slide, extracts per-patch embeddings, and aggregates them into a slide-level representation. Multiple instance learning (MIL) is the dominant weakly supervised paradigm, treating each slide as a bag of patches with a single slide-level label. Attention-based MIL (ABMIL) [2] and CLAM [3] are the canonical aggregators, extended by transformer [4–7], dual-stream [8] and pseudo-bag distillation [9] variants, by graph models that encode tissue architecture [10, 11], and, for gastric cancer, by recalibrated and hybrid convolutional neural network (CNN)–transformer models [12, 13] and by multiple-bag MIL with foundation-model tumour detection for Epstein–Barr virus status prediction in multicentre biopsies [14]. ACMIL [15] combines multi-branch attention with stochastic top-instance masking and a diversity penalty. In parallel, self-supervised and foundation-model pretraining has sharply improved the patch embeddings themselves, from hierarchical pretraining [16] to general-purpose models such as UNI [17], Virchow [18] and GenBio-PathFM [19], and recent multi-centre work indicates that frozen foundation-model features, not aggregator design, increasingly determine downstream MIL accuracy [20].

The barrier to clinical deployment is cross-hospital distribution shift. Staining, fixation, tissue processing and scanner hardware shift patch-feature distributions even for the same tissue type [21], so models trained at one site degrade on slides from a new hospital, especially when 20× and 40× scanners coexist. Training-time remedies include stain normalisation and augmentation [22, 23] and domain-adversarial training [24–26], but they need the multi-site variation, and often domain labels, during training and couple robustness to a trainable backbone, so onboarding a new scanner means retraining. Pixel-level stain normalisation [27–29] requires re-extracting features for every site, whereas feature-level alignment such as correlation alignment (CORAL) [30] and contrastive domain adaptation [31] adapt the learned representation instead. Two further gaps limit trust. Attention maps and gradient attributions [32] are usually judged for plausibility rather than for whether the highlighted regions actually drive the decision, and calibrated uncertainty is increasingly regarded as a prerequisite for clinical AI [33]. Most MIL systems are also developed and validated within a single institution, leaving their cross-hospital behaviour uncharacterised.

Here we measure where MIL accuracy comes from under cross-hospital shift, and what a system must expose to be deployed across hospitals. We assembled a five-hospital leave-one-site-out (LOSO) benchmark of 5,036 H&E-stained gastric slides from five Korean tertiary centres scanned at 20× and 40×, and evaluated a hierarchical patch–region–slide graph aggregator (CHIME-MIL) against eight published aggregators and global max pooling on frozen UNI and GenBio-PathFM embeddings [17, 19]. We tested a label-free, first-moment alignment that subtracts each hospital's mean embedding at inference using only the scanner identifier, including a strict variant that uses no held-out-hospital information and an onboarding curve for new hospitals, and we tested calibration and the faithfulness of the attention maps. The central finding is a negative one for architecture: with strong foundation-model features, the hierarchical aggregator does not exceed attention-based pooling under a matched training recipe, and global max pooling given the same recipe is significantly better ($P = 0.0002$), so the value of an aggregator lies in label-free onboarding, calibration checks and auditable evidence rather than in accuracy.

## Results

### Study design

The benchmark comprises 5,036 H&E-stained gastric WSIs from five Korean tertiary hospitals (Sites A–E; Table 1), scanned at 20× or 40×. Every model was evaluated by leave-one-site-out (LOSO) cross-validation, in which each hospital in turn is held out for testing and never used for training or model selection. Figure 1 summarises the pipeline. Frozen GenBio-PathFM patch embeddings are aligned by subtracting a per-site mean vector, processed by a three-level hierarchical encoder (patch, region, slide graph), and the predictions of two models trained at 256 and 392 px are averaged (Methods).

**Table 1.** Per-hospital composition of the five-site dataset. Each hospital forms one held-out fold in the LOSO evaluation. Mag. = scanner objective; MPP = native microns-per-pixel (`openslide.mpp-x`), used for resolution normalisation; M = malignant; N = normal. †Site B raw slides are archived off-site; the MPP is the 40× scanner-class nominal, consistent with the measured Site E value.

| Site | Mag. | MPP | M | N | Total | Stain |
|---|---|---|---|---|---|---|
| Site A | 20× | 0.50 | 811 | 1,825 | 2,636 | H&E |
| Site B | 40× | 0.25† | 344 | 839 | 1,183 | H&E |
| Site C | 20× | 0.50 | 125 | 209 | 334 | H&E |
| Site D | 20× | 0.50 | 337 | 348 | 685 | H&E |
| Site E | 40× | 0.25 | 57 | 141 | 198 | H&E |
| **Total** | | | **1,674** | **3,362** | **5,036** | |

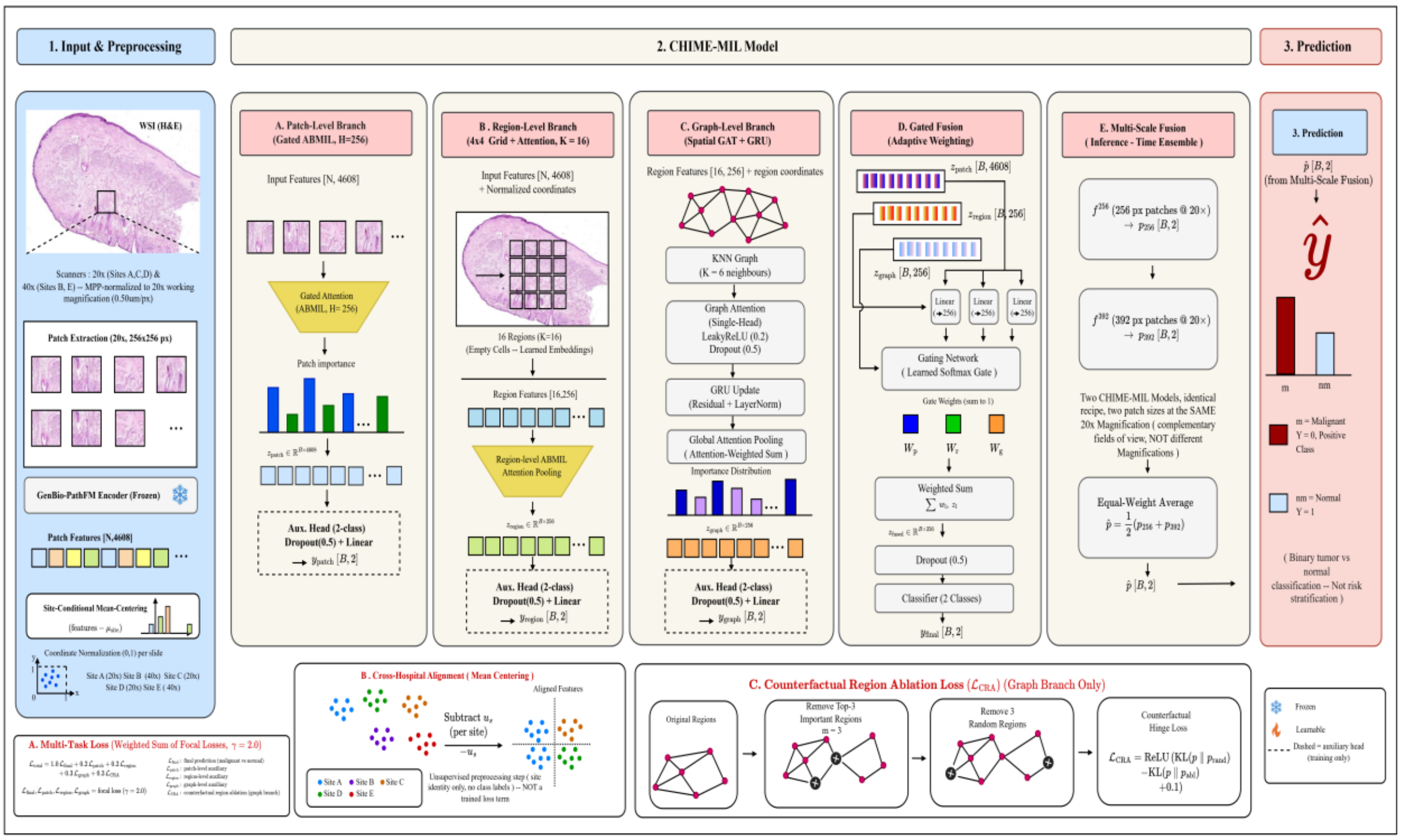

**Figure 1.** Overview of the CHIME-MIL pipeline. (**1**) Input WSIs are tiled at 20×, encoded by the frozen GenBio-PathFM encoder, and site-conditionally mean-centred. (**2**) The hierarchical model processes patches through three parallel branches: (A) gated ABMIL for patch-level importance, (B) a 4×4 grid aggregator for region-level features, and (C) a spatial graph attention network (GAT) with a gated recurrent unit (GRU) update for graph-level reasoning, combined by adaptive gated fusion (D); two models trained at 256 and 392 px are ensembled by equal-weight averaging (E). (**3**) The fused embedding is classified as malignant ($Y = 0$, positive class) or normal ($Y = 1$). (**4**) Training uses a multi-task focal loss ($\gamma = 2.0$) at each level plus a counterfactual region ablation hinge loss.

## Aggregator complexity does not improve discrimination

Table 2 reports five-fold LOSO performance. On the UNI track, ABMIL is the strongest attention-MIL baseline (0.8199). The headline model, CHIME-MIL with GenBio-PathFM features and multi-scale fusion, reaches an area under the receiver operating characteristic curve (AUC) of 0.8374 (±0.0041 over three seeds; deterministic seed-42 reference 0.8434). Against the six conventional GenBio baselines, the paired bootstrap (Methods) favours the headline on five-fold-mean AUC for all six (ΔAUC +0.0100 to +0.0301, Benjamini–Hochberg-adjusted $P_{BH} \leq 0.018$; ABMIL +0.0102, $P_{BH}=0.014$). On the pre-specified pooled-LOSO AUC it exceeds five of them (+0.0075 to +0.0363, $P_{BH} \leq 0.032$) but not ABMIL (−0.0024, $P_{BH}=0.44$); the pooled estimate is dominated by Site A (52% of slides) and also depends on how comparable the score scales are across sites. Against the foundation-model-ceiling probes the direction reverses. MaxPool at 256 and 392 px and ACMIL all exceed the headline on pooled AUC (ΔAUC −0.0176, −0.0153, −0.0093; $P_{BH}<0.001$, 0.0003, 0.0042), and are numerically higher but not significantly different on fold-mean AUC (−0.0055, −0.0087, −0.0012; $P_{BH}=0.36$, 0.11, 0.77).

These comparisons set the full headline recipe (site alignment, hospital-stratified validation and two-scale fusion) against single-scale, unaligned baselines, so they compare recipes rather than aggregators. Under a matched recipe at 256 px the hierarchical aggregator does not exceed ABMIL: without alignment the two reach 0.8204 and 0.8272 (ΔAUC=−0.0068, 95% confidence interval (CI) −0.0155 to +0.0019, $P=0.12$; pooled −0.0068, $P=0.026$), and with mean-centring 0.8274 and 0.8358 (−0.0084, −0.0175 to +0.0010, $P=0.078$; pooled +0.0045, $P=0.19$). Given the same mean-centring and 256+392 px fusion as the headline, global max pooling reaches 0.8552 and exceeds the three-seed headline significantly (−0.0178, 95% CI −0.0261 to −0.0092, $P=0.0002$; pooled −0.0262, $P<0.0002$), although it lacks stratified validation (Table 3). The choice of aggregator does matter, but in favour of simplicity: with every baseline under the same single-scale, unaligned recipe, max pooling at 256 px exceeds ABMIL (+0.0157), DSMIL (+0.0264), RoFormer (+0.0305) and TransMIL (+0.0356) on five-fold-mean AUC ($P \leq 0.0004$ for each, also after Benjamini–Hochberg correction across the four tests; pooled AUC +0.015 to +0.054, all $P<0.0002$). We therefore claim no discrimination advantage for the hierarchical architecture: with strong patch features, hierarchical complexity does not improve AUC over simpler aggregators on this binary task.

**Table 2.** Five-fold LOSO comparison. The **AUC** column is the five-fold mean (mean ± std over held-out sites). $\boldsymbol{P_{BH}}$: paired slide-level bootstrap ($B$=10,000, seed 42, two-sided) of the CHIME-MIL headline against each GenBio-PathFM baseline, with the headline statistic taken as the mean over its three training seeds, so that the test concerns the tabulated headline. The first value tests the tabulated five-fold-mean AUC (slides resampled within each held-out site); the second tests the pre-specified pooled-LOSO AUC ($N$=5,036; headline pooled AUC 0.8143). $P$ values are Benjamini–Hochberg-adjusted within the two pre-specified families (see the *Evaluation protocol and statistics* subsection of the Methods): the six conventional baselines, and the three foundation-model-ceiling probes (ACMIL, MaxPool 256/392 px). < 0.001 means no bootstrap replicate crossed zero. Signed differences are given in the *Aggregator complexity does not improve discrimination* subsection of the Results. UNI-track significance is not asserted (per-slide predictions not retained). ‡ The headline AUC is the three-seed mean over {42,123,7} (0.8374 ± 0.0041 across seeds; the ±0.028 shown is the across-site std, as in every row), using the label-free all-slides $\boldsymbol{\mu}_s$ ("leaky" only in the unsupervised test-time sense; no labels enter it). A strict-LOSO variant that estimates $\boldsymbol{\mu}_s$ from training sites only costs ΔAUC=−0.0019 (Supplementary Table S4). The deterministic seed-42 reference run reproduced by the released code reaches 0.8434. Baseline GenBio rows are single-scale, *without* first-moment alignment or hospital-stratified validation, so these tests compare the full headline recipe with each baseline's recipe; alignment is evaluated in Table 3 and recipe-matched comparisons are given in the Results. MC = site mean-centring (first-moment alignment). Bold = best per column.

| Method | AUC | ACC | BalAcc | $P_{BH}$ fold-mean / pooled |
|---|---|---|---|---|
| *Baselines (UNI features, 1,024-d)* | | | | |
| ABMIL [2] | 0.8199 ± 0.040 | 0.7546 | – | – |
| CLAM-SB [3] | 0.8124 ± 0.036 | 0.6744 | – | – |
| CLAM-MB [3] | 0.8164 ± 0.049 | 0.7339 | – | – |
| TransMIL [4] | 0.8095 ± 0.046 | 0.7202 | 0.7214 | – |
| DSMIL [8] | 0.7657 ± 0.040 | 0.6432 | 0.6830 | – |
| DTFD-MIL [9] | 0.8158 ± 0.035 | 0.7181 | 0.7351 | – |
| *Baselines (GenBio-PathFM features, 4,608-d)* | | | | |
| ABMIL [2] | 0.8272 ± 0.032 | 0.7504 | 0.7583 | 0.014 / 0.44 |
| CLAM-SB [3] | 0.8208 ± 0.033 | 0.7553 | 0.7561 | 0.0006 / 0.015 |
| CLAM-MB [3] | 0.8274 ± 0.025 | 0.7544 | 0.7528 | 0.018 / 0.032 |
| TransMIL [4] | 0.8073 ± 0.046 | 0.7440 | 0.7541 | < 0.001 / < 0.001 |
| DSMIL [8] | 0.8165 ± 0.037 | 0.7597 | 0.7477 | < 0.001 / < 0.001 |
| RoFormer [5] | 0.8125 ± 0.031 | 0.7357 | 0.7497 | < 0.001 / < 0.001 |
| ACMIL [15] | 0.8386 ± 0.020 | 0.7602 | 0.7627 | 0.77 / 0.0042 |
| ACMIL + MC [15] | 0.8370 ± 0.025 | 0.7336 | 0.7496 | – |
| MaxPool (256 px) | 0.8429 ± 0.024 | 0.7549 | 0.7459 | 0.36 / < 0.001 |
| MaxPool (392 px) | **0.8461 ± 0.029** | 0.7680 | 0.7613 | 0.11 / 0.0003 |
| *CHIME-MIL (ours): UNI comparison track* | | | | |
| CHIME-MIL (single model) | 0.8257 | 0.7467 | – | – |
| CHIME-MIL (Top-4 ensemble) | 0.8346 ± 0.039 | 0.7697 | 0.7554 | – |
| *CHIME-MIL (ours): GenBio headline* | | | | |
| CHIME-MIL (mc, 256 + 392 px) | 0.8374 ± 0.028‡ | **0.7729 ± 0.036** | **0.7683 ± 0.032** | – |

**Table 3.** Ablation study. All CHIME-MIL rows use GenBio-PathFM features (4,608-d) with the spatial-grid region aggregator, except the first two rows which use UNI (1,024-d) as the comparison reference. MC = first-moment site alignment; SV = hospital-stratified validation split; MS = multi-scale late fusion (256+392 px). The counterfactual region ablation loss is active ($\lambda_c$=0.3) in every row except the explicitly marked "no CRA" row, which sets $\lambda_c$=0 under an otherwise identical base configuration, isolating the loss as a controlled A/B against the base row. Mean AUC ± std over five LOSO folds. CHIME-MIL GenBio rows (base, +MC, +MC+SV, full, MC+SV @ 256+392) are the deterministic seed-42 reference run (fixed val split, `CUBLAS_WORKSPACE_CONFIG`, `PYTHONHASHSEED`=42) that the released code reproduces exactly, used here for reproducible component attribution (it removes run-to-run GPU non-determinism, not seed variance; paired bootstrap tests for each step are given in the text); the headline performance reported in Table 2 is the three-seed mean 0.8374 ± 0.0041 (seeds 42, 123 and 7 give 0.8332, 0.8377 and 0.8414); the deterministic reference run (0.8434) is a separate re-execution. Rows marked † (no CRA, MC only @ 256+392, and the MaxPool control block) are from the earlier runs and are shown for reference. The MaxPool + MC @ 256+392 row is the equal-weight probability average of the two MaxPool+MC rows above it, i.e. the same late-fusion rule as the headline, and is the recipe-matched comparison taken up in the Discussion (balanced accuracy 0.7756; AUC above the three-seed headline, $P$=0.0002, paired bootstrap). MaxPool and ABMIL rows use a label-stratified random validation split (SV does not apply).

| Configuration | Feat. | MC | SV | MS | AUC |
|---|---|---|---|---|---|
| ABMIL (reference) | UNI | – | – | – | 0.8199 ± 0.040 |
| CHIME-MIL, no alignment | UNI | ✗ | ✗ | ✗ | 0.8257 |

| Configuration | Feat. | MC | SV | MS | AUC |
|---|---|---|---|---|---|
| CHIME-MIL, no alignment | GenBio | ✗ | ✗ | ✗ | 0.8204 ± 0.035 |
| no CRA ($\lambda_c$=0) | GenBio | ✗ | ✗ | ✗ | 0.8213 ± 0.030† |
| + MC | GenBio | ✓ | ✗ | ✗ | 0.8274 ± 0.029 |
| + MC + SV | GenBio | ✓ | ✓ | ✗ | 0.8341 ± 0.027 |
| + MC + SV + MS (full) | GenBio | ✓ | ✓ | ✓ | **0.8434 ± 0.030** |
| *Multi-scale fusion cross comparison* | | | | | |
| MC only @ (256+392) | GenBio | ✓ | ✗ | ✓ | 0.8395† |
| MC + SV @ (256+392) | GenBio | ✓ | ✓ | ✓ | **0.8434** |
| *MaxPool control: MC on a non-hierarchical aggregator* | | | | | |
| MaxPool 256 px | GenBio | ✗ | — | ✗ | 0.8429 ± 0.024† |
| MC | GenBio | ✓ | — | ✗ | 0.8510 ± 0.028† |
| MaxPool 392 px | GenBio | ✗ | — | ✗ | 0.8461 ± 0.029† |
| MC | GenBio | ✓ | — | ✗ | 0.8488 ± 0.030† |
| MC @ (256+392) | GenBio | ✓ | — | ✓ | 0.8552 ± 0.029† |
| *ABMIL control: MC on a gated-attention aggregator* | | | | | |
| ABMIL 256 px | GenBio | ✗ | — | ✗ | 0.8272 ± 0.032 |
| MC | GenBio | ✓ | — | ✗ | 0.8358 ± 0.031 |
| ABMIL 392 px | GenBio | ✗ | — | ✗ | 0.8269 ± 0.029 |
| MC | GenBio | ✓ | — | ✗ | 0.8346 ± 0.034 |

## Per-site performance

Table 4 breaks AUC down by hospital. Site C shows the largest alignment gain: mean-centring alone raises its AUC from 0.7835 to 0.8060 (+0.0225, 95% CI −0.0009 to +0.0471, $P$=0.061; paired bootstrap within the site), in line with a feature-space diagnostic identifying Site C normal tissue as closest to the tumour cluster (Supplementary Note 2). Mean-centring raises AUC at four of the five sites and lowers it at Site E (0.8601 to 0.8503; Supplementary Table S2). Site A is the hardest site for almost every method despite its size (the exceptions, TransMIL on both feature tracks and DSMIL on GenBio-PathFM, do worse on Site C): in an audit of four earlier CHIME-MIL variants (with and without mean-centring), 555 of its slides (21.1% of Site A) were misclassified by all four at a 0.5 threshold, so much of the Site A error is shared across model variants rather than specific to one configuration. CHIME-MIL and ACMIL are the strongest per-site attention aggregators; they show no significant difference on fold-mean AUC ($P_{BH}$=0.77), while ACMIL is higher on pooled AUC (ΔAUC=−0.0093, $P_{BH}$=0.0042; Table 2).

**Table 4.** Per-site AUC under LOSO evaluation. Each column is the AUC on one held-out hospital. CHIME-MIL (UNI) is the Top-4 ensemble; CHIME-MIL (GenBio) is the mc multi-scale (256+392 px) deterministic seed-42 reference run (its per-site values average to 0.8434; the three-seed headline mean is 0.8374, Table 2). Bold = best result per site across all methods.

| Method | Site A | Site B | Site C | Site D | Site E |
|---|---|---|---|---|---|
| *Baselines (UNI features, 1,024-d)* | | | | | |
| ABMIL | 0.766 | 0.845 | 0.797 | 0.811 | 0.880 |
| CLAM-SB | 0.769 | 0.850 | 0.785 | 0.799 | 0.859 |

| Method | Site A | Site B | Site C | Site D | Site E |
|---|---|---|---|---|---|
| CLAM-MB | 0.763 | 0.849 | 0.772 | 0.805 | **0.893** |
| TransMIL | 0.758 | 0.848 | 0.755 | 0.819 | 0.868 |
| DSMIL | 0.727 | 0.768 | 0.729 | 0.769 | 0.836 |
| DTFD-MIL | 0.769 | 0.838 | 0.790 | 0.814 | 0.868 |
| *Baselines (GenBio-PathFM features, 4,608-d)* | | | | | |
| ABMIL | 0.791 | 0.855 | 0.797 | 0.820 | 0.873 |
| CLAM-SB | 0.785 | 0.848 | 0.790 | 0.811 | 0.870 |
| CLAM-MB | 0.791 | 0.852 | 0.806 | 0.834 | 0.854 |
| TransMIL | 0.769 | 0.825 | 0.748 | 0.815 | 0.880 |
| DSMIL | 0.784 | 0.824 | 0.775 | 0.822 | 0.878 |
| RoFormer | 0.766 | 0.846 | 0.797 | 0.807 | 0.847 |
| ACMIL | 0.804 | 0.859 | **0.836** | 0.836 | 0.859 |
| MaxPool (256 px) | **0.807** | 0.862 | 0.823 | **0.853** | 0.870 |
| MaxPool (392 px) | 0.803 | 0.863 | 0.831 | 0.845 | 0.889 |
| *CHIME-MIL (ours)* | | | | | |
| CHIME-MIL (UNI, Top-4 ens.) | 0.785 | 0.855 | 0.816 | 0.829 | 0.888 |
| CHIME-MIL (GenBio, mc) | 0.793 | **0.870** | 0.835 | 0.840 | 0.878 |

## Component ablation and alignment transfer

Table 3 traces each component (deterministic seed-42 run); each step is tested with the paired bootstrap on five-fold-mean AUC (Methods; 95% CI, unadjusted $P$). Multi-scale fusion is the only step with a statistically supported gain on CHIME-MIL: +0.0092 (95% CI +0.0038 to +0.0149, $P$=0.001), reaching 0.8434, with positive per-fold gains everywhere (Supplementary Table S2). Mean-centring (MC) raises AUC from 0.8204 to 0.8274 (+0.0070, 95% CI −0.0015 to +0.0157, $P$=0.10), concentrated at Site C, and hospital-stratified validation (SV) adds +0.0067 (−0.0034 to +0.0171, $P$=0.20): both are positive but neither is significant on its own in a single seed. The counterfactual region ablation (CRA) loss does not help: at the headline configuration (MC+SV, 256 px) disabling it ($\lambda_c$=0) changes fold-mean AUC by +0.0058 (−0.0022 to +0.0137, $P$=0.16) and raises pooled AUC by +0.0155 ($P$<0.0002), with no loss of attention concentration (the earlier toggle without alignment in Table 3 gives +0.0009). We keep $\lambda_c$=0.3 only for specification fidelity and report it as a documented design choice (Supplementary Note 7). The accuracy effect of alignment on other aggregators depends on the statistic. On five-fold-mean AUC mean-centring lifts MaxPool (+0.0081, 95% CI +0.0019 to +0.0143, unadjusted $P$=0.009) and ABMIL (+0.0086, +0.0003 to +0.0166, $P$=0.044) at least as much as CHIME-MIL (+0.0070), but after Benjamini–Hochberg correction across the three transfer tests (MaxPool, ABMIL, ACMIL) only MaxPool remains significant ($P_{BH}$=0.026 and 0.065). On pooled AUC the gains disappear: MaxPool +0.0042 ($P$=0.14), CHIME-MIL +0.0037 ($P$=0.31), and ABMIL falls by 0.0077 ($P$=0.048, $P_{BH}$=0.14). At 256 px MaxPool+MC (0.8510) exceeds CHIME-MIL+MC (0.8274). Alignment does not help ACMIL on either statistic (0.8386 → 0.8370, −0.0016, $P$=0.70), and disabling ACMIL's top-instance masking does not change this (−0.0012), so the non-transfer attaches to its diversity-regularised multi-branch attention (Supplementary Note 10).

### First-moment alignment removes most of the hospital signal

Raw per-slide embeddings segregate by hospital, with the two 40× scanners (Sites B and E) clearly separated (Fig. 2A). After mean-centring the hospital clusters dissolve (Fig. 2B): a 15-nearest-neighbour hospital classifier (majority-class baseline 0.52) falls from 0.910 to 0.709 accuracy and the linear inter-site maximum mean discrepancy (MMD) from 35.96 to 0.54. Second-moment alternatives do not help (Supplementary Table S3): CORAL (0.8144) and per-site zero-phase component analysis (ZCA) whitening (0.8248) both trail mean-centring (0.8274) on CHIME-MIL, and similarly for ABMIL. Whitening strips more site structure (residual 15-NN 0.587 and 0.570) at $\mathcal{O}(d^2)$ cost yet yields no gain, so the residual covariance is site-identifying but task-irrelevant. A trivial per-slide centring is worse than no alignment (0.8173), so the site, not the slide, is the correct granularity.

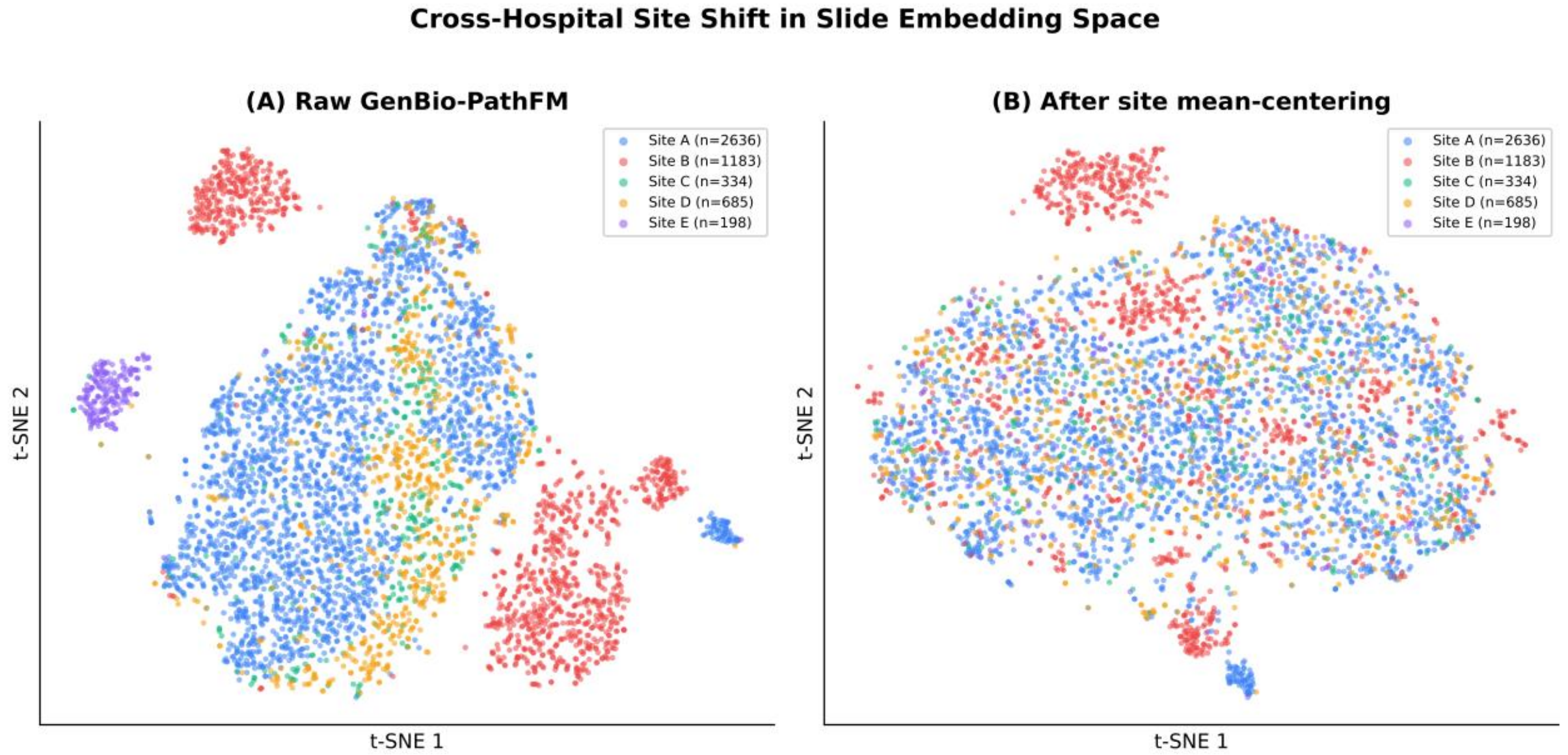

**Figure 2.** Cross-hospital site shift in slide-embedding space. t-distributed stochastic neighbour embedding (t-SNE) of per-slide GenBio-PathFM embeddings. **A**, Raw: slides segregate by hospital, the two 40× scanners (Sites B, E) clearly separated. **B**, After site mean-centring: per-site clusters dissolve and slides intermix. An inter-site 15-nearest-neighbour hospital classifier (majority-class baseline 0.52) falls from 0.910 to 0.709 accuracy and the linear inter-site MMD from 35.96 to 0.54; the residual reflects higher-moment structure (Supplementary Table S3).

### The headline does not depend on held-out-hospital statistics

Replacing the held-out-site mean with the average of the four training-site means, which uses no held-out information, costs only ΔAUC=−0.0019 on the fused model (0.8374 to 0.8355; Supplementary Table S4). Figure 3 sweeps the number of unlabelled onboarding slides $N$ for the single-scale 256 px model (the seed-42 run of the three-seed sweep; Methods): across folds the mean AUC moves by less than $10^{-3}$ between $N$=10 and $N$=all, so a new hospital is onboarded from $N^{*}$=10 unlabelled slides with no retraining.

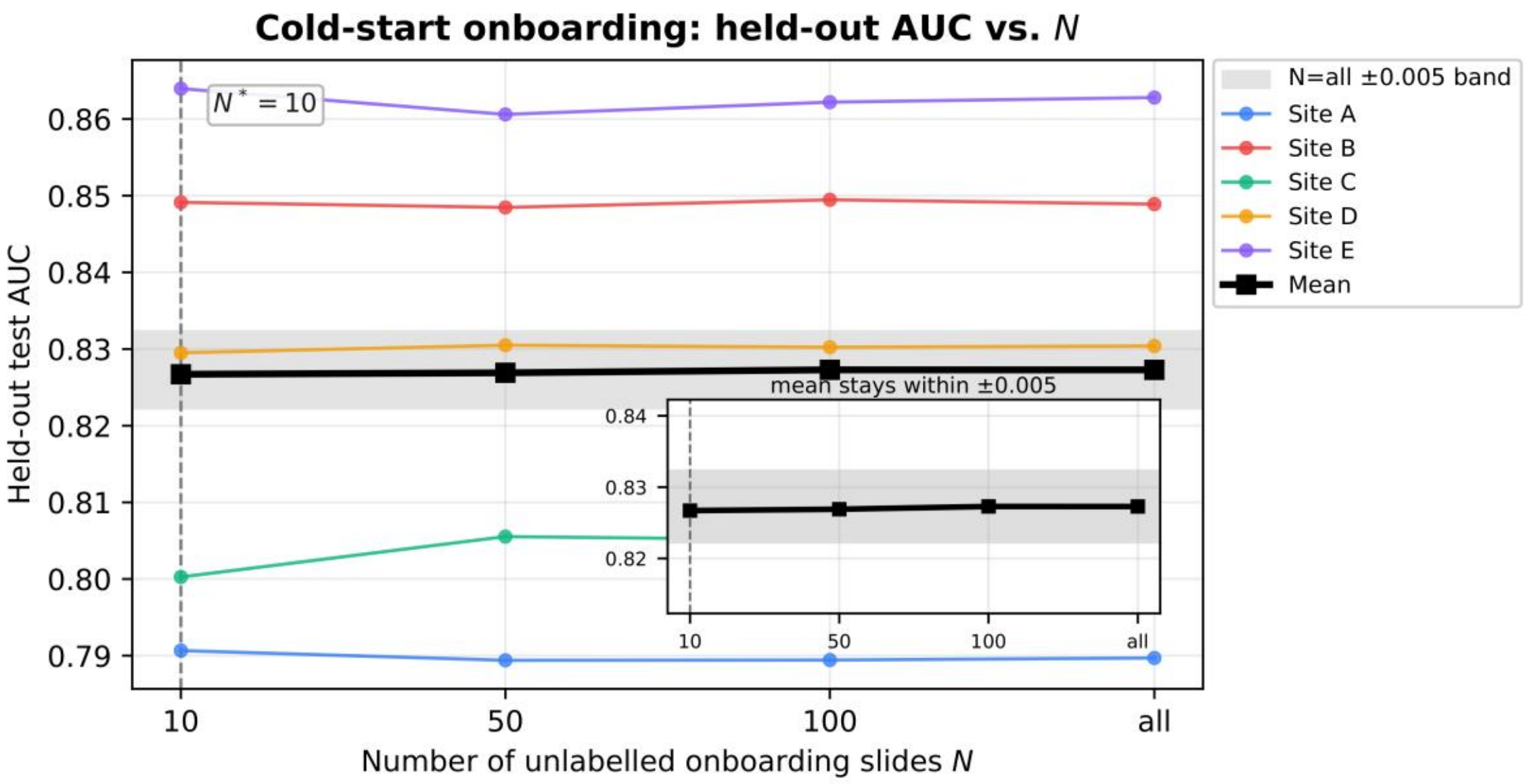

**Figure 3.** Cold-start onboarding curve. Held-out AUC versus the number of unlabelled onboarding slides $N \in \{10,50,100,\text{all}\}$ used to estimate $\boldsymbol{\mu}_s$ at each held-out hospital (no retraining; single-scale 256 px model, seed-42 run of the three-seed sweep). Coloured lines show per-site AUC (error bars over three subset-sampling seeds); the black line is the across-folds mean and the grey band the ±0.005 AUC tolerance around the $N$=all ceiling (0.8273). By $N^{*}$=10 all curves lie within the band.

## External validation on CAMELYON16

Applying the unchanged pipeline to the public CAMELYON16 benchmark [34] (the single-source Radboud subset: 246 sentinel lymph-node WSIs, breast-cancer metastasis), with the same frozen GenBio-PathFM encoder and five-fold stratified cross-validation (five training seeds for CHIME-MIL, one for each baseline), CHIME-MIL attains an AUC of 0.9916 ± 0.012 (± across seeds), alongside ABMIL (0.9970) and CLAM-SB (0.9947) (Supplementary Table S5). All three aggregators saturate on this cohort, and a paired slide-level bootstrap over the pooled out-of-fold predictions separates none of them: against ABMIL ΔAUC = − 0.0032 (95% CI −0.0068 to +0.0004, $P_{\mathrm{BH}}$ = 0.15); against CLAM-SB ΔAUC = + 0.0025 (−0.0041 to +0.0130, $P_{\mathrm{BH}}$ = 0.64) ($B$ = 10,000, paired on the same resampled slides; the test uses pooled out-of-fold AUC averaged over seeds, whereas the table reports fold-mean AUC, on which CHIME-MIL is 0.0054 below ABMIL and 0.0031 below CLAM-SB). The benchmark therefore confirms architecture transfer without discriminating between aggregators, the same foundation-model-ceiling pattern the gastric benchmark shows. Because this subset is single-source, the per-site alignment is inactive by design, so this is an architecture-transfer check rather than a test of the alignment.

### Calibration and efficiency

The headline model has a low top-label expected calibration error (ECE 0.053 ± 0.014; Brier score 0.163 ± 0.020), but its malignancy probabilities are systematically under-confident (mean positive-class reliability gap 0.108 ± 0.028). The deviation is monotone, so a post-hoc temperature fitted on validation data could reduce it without changing AUC; we did not fit one. It has 12.29 M parameters (6.15 M per tower) and completes end-to-end aggregation on an H100 GPU at batch size 1 in ≈3.7 ms per slide (for a representative bag of 2,934 patches; the corpus median is ≈3,350), on par with TransMIL; feature extraction dominates the total cost (Supplementary Note 9).

### Attention maps identify the evidence behind each prediction

Figure 4 overlays patch-level gated attention on representative malignant and normal WSIs. Beyond visual concentration, we tested faithfulness with comprehensiveness and sufficiency [35] on all 1,674 malignant slides across the five folds (inference only). Removing the top-attention patches lowers the malignancy probability by 0.104 on average, against 0.0006 for the same number of random patches (paired difference 0.103, 95% CI 0.097 to 0.110; larger for attention on 80% of slides; two-sided Wilcoxon signed-rank test, $n$=1,674, $P$=2.7 × $10^{-148}$), and retaining only the top-attention patches recovers the full-slide prediction (sufficiency = − 0.035). The attended patches therefore carry the predictive evidence rather than merely looking plausible, which is the property a pathologist-facing audit trail requires (population statistics in Supplementary Note 6).

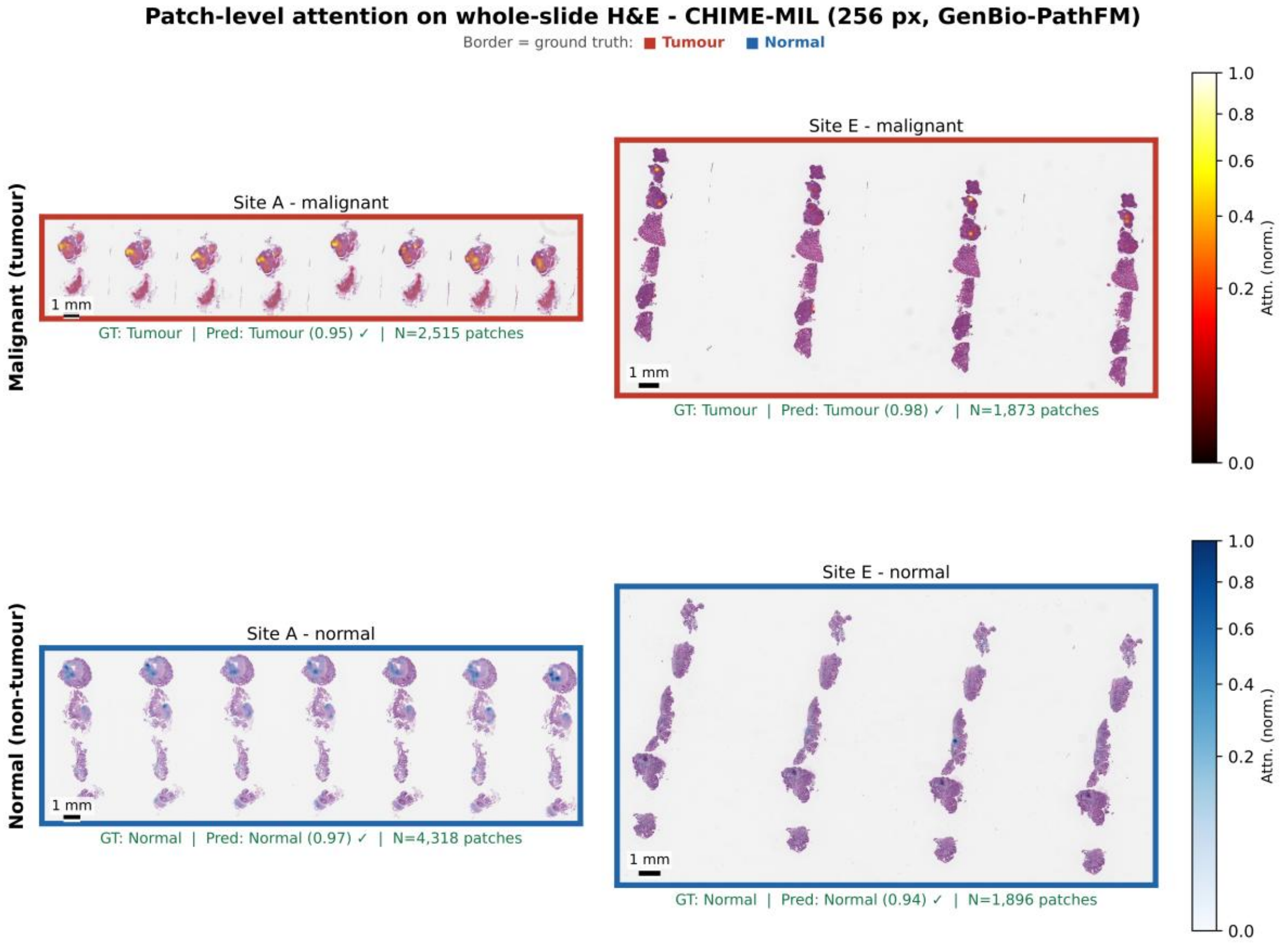

**Figure 4.** Patch-level attention on representative whole-slide images. The normalised gated-attention weight $\tilde{\alpha}_i \in [0,1]$ from the patch MIL head (GenBio-PathFM, mean-centred, 256 px) is rendered as a heatmap on the slide thumbnail for WSIs from Site A and Site E. Top row: malignant WSIs correctly classified as tumour; bottom row: normal WSIs correctly classified as non-tumour. Panel borders encode ground truth (red, tumour; blue, normal); ground-truth label, prediction, confidence and patch count are shown below each panel, with a 1 mm scale bar from each slide's `openslide.mpp-x`. A check mark beside the prediction denotes agreement with the ground-truth label and a cross denotes disagreement.

## Discussion

The central finding is that, with strong foundation-model features, aggregator complexity is not what drives discrimination on this binary task. Read recipe-for-recipe, when global max pooling is given the same first-moment alignment and 256+392 px fusion as the headline model, it reaches an AUC of 0.8552 and a balanced accuracy of 0.7756, above CHIME-MIL on both (Table 3); the AUC difference is significant (ΔAUC=−0.0178, 95% CI −0.0261 to −0.0092, *P*=0.0002, paired bootstrap against the three-seed headline) even though max pooling omits stratified validation, and under a matched recipe the hierarchical aggregator does not exceed ABMIL either (Results). A leakage-free decomposition shows that the fused prediction is not significantly different from its own patch (gated-ABMIL) branch on fold-mean AUC (ΔAUC=+0.0010, 95% CI −0.0084 to +0.0104, *P*=0.84, paired bootstrap) and is lower on pooled AUC (−0.0059, *P*=0.025), with the three branches mutually correlated at 0.92–0.97 (Supplementary Note 10). Within CHIME-MIL, aggregator-side engineering (gating, graph topology, diversity penalties) recovers at most a few thousandths of AUC; a substantial gain would require intervening on the feature side. This ceiling agrees with recent multi-centre evidence that frozen foundation-model features, not aggregator design, increasingly determine downstream MIL accuracy [20]. The comparative advantage of a hierarchical aggregator is therefore not raw discrimination but the patch- and region-level attention that pooling cannot expose, together with label-free cross-site deployment.

The first-moment alignment is the deployable part of this result. It needs no class labels, only the scanner identifier that every deployment already records; it costs only 0.0019 AUC when the held-out hospital contributes nothing to the site mean; and a new hospital is onboarded from about ten unlabelled slides without retraining, which turns the recurring cost of a multi-site rollout into a one-time mean computation. Its accuracy effect, by contrast, is small and depends on the statistic: on hospital-averaged AUC it helped max pooling ($P$=0.009; $P_{BH}$=0.026) and, less clearly, ABMIL ($P$=0.044; $P_{BH}$=0.065), was not significant on CHIME-MIL ($P$=0.10) and absent on ACMIL, and on pooled AUC no aggregator gained (ABMIL lost 0.0077, $P$=0.048). The case for the alignment is therefore deployability (no labels, no retraining, negligible cost) rather than accuracy. Higher-order alignment removed more hospital signal but did not improve accuracy, so the residual covariance identifies the hospital without carrying task information.

The counterfactual region ablation loss should be read as an attention-concentration regulariser, not a structural causal model: zeroing top-attention nodes is a region ablation heuristic and performs no causal identification. It improved neither AUC nor attention concentration, and disabling it raised pooled AUC (Supplementary Note 7); we retain $\lambda_c$=0.3 only to preserve the original specification and report it as a documented design choice. Its ablation count is not inert, however: in a single sweep, ablating $m$=5 regions gave a significantly higher AUC than the default $m$=3 (+0.016, $P_{BH}$<0.001, paired bootstrap; Supplementary Note 5). We did not adopt it, because choosing $m$ from these results would tune on the held-out hospitals.

The residual errors are morphological rather than distributional. First-moment centring corrects the dominant distributional shift, but a cell-level audit indicates that the residual false positives on Sites C and A track a stromal and architectural composition difference (normal tissue resembling tumour) rather than tumour-cell content or a removable scanner offset, and this replicates on a second held-out site (Supplementary Note 3). A direct anti-stroma attention penalty leaves discrimination intact but does not give a seed-robust reduction in these false positives (Supplementary Note 4), so they remain an open problem.

This study has limitations. The benchmark covers a single binary task (malignant versus normal) on gastric tissue from five Korean tertiary centres, so multi-class subtyping and inter-continental population variation are not represented. Patient sex and gender were not available in the de-identified metadata and were not analysed. All experiments are retrospective, and the decision-support framing rests on the premise, not yet tested in a reader study, that the surfaced attention is actionable for pathologists; the interpretability claim should be read as auditability of the surfaced evidence, not as validated diagnostic utility. The external CAMELYON16 result supports architecture transfer only, because its single-source data disables the per-site alignment by construction. A faithful external test of the alignment requires a public corpus that is both multi-institution and labelled for tumour versus normal; the natural candidates, multi-site CAMELYON17 (binary but breast lymph node) and organ-matched TCGA-STAD (gastric but tumour-heavy), each meet only part of this.

Future work includes adding the native 40× scale and joint two-tower training, upgrading or site-invariantly fine-tuning the feature backbone, validating the alignment on public multi-site cohorts such as CAMELYON17 and TCGA/CPTAC, and extending from binary classification to multi-class gastric-cancer subtyping. In summary, on a five-hospital benchmark with strong foundation-model features, the hierarchical aggregator does not improve discrimination over simpler aggregators, global max pooling included, and its value lies in label-free onboarding, calibration checks and auditable per-slide evidence.

## Methods

### Cohort and ethics approval

The dataset comprises 5,036 H&E-stained gastric WSIs collected retrospectively from five tertiary care hospitals in the Republic of Korea (Sites A–E), spanning two scanner magnifications (20×, 40×) and distinct staining and processing protocols. Slides come from endoscopic biopsy and surgical resection specimens; the task is slide-level classification of gastric carcinoma (predominantly adenocarcinoma) versus non-tumour mucosa, with binary labels and no patch-level annotations. Table 1 gives the per-hospital composition: Site A is largest (2,636 slides, 52.3%), Site E smallest (198, 3.9%), and most sites are class-imbalanced (≈1:2 malignant:normal). Tumour slides have larger tissue areas than normal slides (median 3,657 versus 3,164 tissue patches of 128 × 128 μm, i.e. 59.9 versus 51.8 $mm^2$; two-sided

Mann–Whitney $U$=3,228,166, $n$=1,674 malignant and 3,362 normal slides, $P$=1.6 × $10^{-17}$, rank-biserial correlation 0.15; the difference has the same direction at every site).

The study was approved by the Institutional Review Boards of the participating institutions, Yonsei University Hospital, Dankook University Hospital, CHA Bundang Medical Center, Ewha Womans University Medical Center (Seoul and Mokdong hospitals) and Keimyung University Dongsan Medical Center (protocol codes SEUMC 2021-07-016, EUMC 2021-08-013, DKUH 2022-03-041, KUDH 2021-08-112, CHABH 2021-09-021-011 and YUMC 4-2021-0029). All methods were carried out in accordance with the relevant guidelines and regulations, including the Declaration of Helsinki. The requirement for informed consent was waived by these Institutional Review Boards because the study was retrospective and used only de-identified archival whole-slide images.

**Problem formulation**

Each WSI from site $s \in \mathcal{S}$ is a bag $\mathcal{X}^{(s)} = \{(\boldsymbol{x}_i, \boldsymbol{c}_i)\}_{i=1}^{N}$ of patch features $\boldsymbol{x}_i \in \mathbb{R}^d$ with normalised coordinates $\mathbf{c}_i \in [0,1]^2$, carrying a single slide-level label $Y \in \{0,1\}$ (patch labels unavailable). Following the underlying label files, $Y$=0 denotes the positive (malignant) class and $Y$=1 the negative (normal) class; all polarity-dependent metrics use `pos_label`=0, so malignant is the positive class in the conventional clinical sense throughout. We learn $f_\theta: \mathcal{X}^{(s)} \mapsto [0,1]$ under leave-one-site-out (LOSO) validation: at fold $k$, all slides of hospital $k$ are held out and the rest train the model. The predicted malignancy probability is

$$\hat{p} = f_\theta(\mathcal{X}^{(s)}) = \Pr(Y = 0 \mid \mathcal{X}^{(s)}). \quad (1)$$

**Feature extraction and magnification normalisation**

Patch features use the frozen GenBio-PathFM foundation model [19]. Each WSI is tiled at 20× (0.50 μm/px) into non-overlapping 256 × 256 patches, each mapped to $\boldsymbol{x}_i \in \mathbb{R}^{4608}$; patches with < 20% foreground tissue are discarded. A second feature set at 392 × 392 px (same magnification) provides a wider receptive field for multi-scale fusion. For scanner-magnification normalisation we work in physical micron space using the per-slide microns-per-pixel value from `openslide.mpp-x`, never the unreliable nominal objective power: each patch is read at a constant physical field of view (128 μm at 256 px) and Lanczos-resampled, so 40× slides are down-sampled to the 20× grid. This removes the resolution component of scanner shift but leaves residual stain, illumination and sensor differences, which are the target of the inference-time alignment below. Features are cached once per magnification (no gradients through the extractor).

**Hierarchical graph encoder**

The encoder processes each bag through three levels before fusing the resulting slide embeddings (Fig. 1).

**Level 1: patch MIL head.**

A gated attention mechanism [2] scores each patch:

$$a_i = \frac{\exp(\boldsymbol{w}^\top(\tanh(\boldsymbol{V}\boldsymbol{x}_i) \odot \sigma(\boldsymbol{U}\boldsymbol{x}_i)))}{\sum_j \exp(\boldsymbol{w}^\top(\tanh(\boldsymbol{V}\boldsymbol{x}_j) \odot \sigma(\boldsymbol{U}\boldsymbol{x}_j)))}, \tag{2}$$

with $\boldsymbol{V}, \boldsymbol{U} \in \mathbb{R}^{H \times d}$, $\boldsymbol{w} \in \mathbb{R}^H$, $H$=256; the patch embedding is $\boldsymbol{z}_{\text{patch}} = \sum_i a_i\, \boldsymbol{x}_i$.

**Level 2: spatial grid region aggregator.**

Each slide is partitioned into a fixed 4×4 grid ($K$=16 regions) over per-slide min–max-normalised coordinates, so the partition is invariant to scanner field of view and slide size by construction. A shared gated-attention head produces per-patch logits $s_i$, softmax-normalised within each cell:

$$\alpha_{ik} = \frac{\exp(s_i)\mathbb{1}[k_i = k]}{\sum_{j:k_j=k} \exp(s_j)}, \qquad \boldsymbol{r}_k = \boldsymbol{W}_r\left(\sum_{i:k_i=k} \alpha_{ik}\, \boldsymbol{x}_i\right) \in \mathbb{R}^{256}. \tag{3}$$

Empty cells take a learned embedding. A region-level attention head pools the $K$ region features into $\boldsymbol{z}_{\text{region}} \in \mathbb{R}^{256}$.

**Level 3: spatial graph reasoning.**

A $\kappa$-nearest-neighbour ($\kappa$=6) graph over region coordinates is built; graph attention over edges $(j,k)$ is

$$e_{jk} = \text{LeakyReLU}\left(\boldsymbol{a}^\top[\boldsymbol{W}_n\boldsymbol{r}_j \parallel \boldsymbol{W}_m\boldsymbol{r}_k]\right), \quad \alpha_{jk} = \text{softmax}_k(e_{jk}), \tag{4}$$

with node states updated by a residual GRU cell over aggregated neighbour messages. A global attention readout pools nodes into $\boldsymbol{z}_{\text{graph}} \in \mathbb{R}^{256}$:

$$\gamma_j = \text{softmax}_j(\boldsymbol{v}_g^\top \tanh(\boldsymbol{W}_g\boldsymbol{h}_j)), \qquad \boldsymbol{z}_{\text{graph}} = \sum_j \gamma_j\, \boldsymbol{h}_j. \tag{5}$$

**Multi-level fusion.**

The three embeddings are projected to a common 256-d space, weighted by a learned gate, and summed:

$$\boldsymbol{z}_{\text{fused}} = \sum_{\ell \in \mathcal{L}} \omega_\ell\, \boldsymbol{W}_\ell \boldsymbol{z}_\ell, \qquad \omega_\ell = \text{softmax}_\ell(\boldsymbol{v}_f^\top \tanh(\boldsymbol{W}_f \boldsymbol{W}_\ell \boldsymbol{z}_\ell)), \tag{6}$$

$\mathcal{L}$ = {patch,region,graph}, with a linear classifier on $\mathbf{z}_{\text{fused}}$ and auxiliary classifiers per level. Headline results use this learned gate (the original specification); a decomposition (Discussion; Supplementary Note 10) shows the branches are highly correlated and a parameter-free uniform gate ($\omega_\ell$=1/3) is a marginally stronger, more seed-robust default.

**Cross-site feature alignment**

Feature distributions vary systematically across sites. A 19.2-million-patch feature-space audit finds that Site C normal tissue has the smallest class-separation cosine margin (Supplementary Note 2), i.e. it lies closest to the tumour cluster, and raw per-slide embeddings segregate by hospital (Fig. 2A). We correct this by subtracting a per-site mean. The alignment uses only the deployment-time site identity $s(i)$ (scanner or picture archiving and communication system metadata), never the class label $Y$. For each site we compute a patch-count-weighted mean over all of that site's slides:

$$\boldsymbol{\mu}_s = \frac{1}{N_s^{\text{tot}}} \sum_{n \in \mathcal{D}_s} \sum_{i=1}^{N_n} \boldsymbol{x}_{n,i}\,, \qquad N_s^{\text{tot}} = \sum_{n \in \mathcal{D}_s} N_n, \tag{7}$$

computed once and shared across folds. Because Equation (7) is a strict function of features and counts (no $Y$), the held-out site contributes only unlabelled features, the standard unsupervised domain-adaptation setting [30, 36]. The relevant criterion is therefore deployment-mirror fidelity rather than the supervised prohibition on held-out data: a strict estimator using only training-site means costs just ΔAUC=−0.0019 (Supplementary Table S4), and a new site is onboarded from 10 unlabelled slides (Fig. 3). At training and inference every patch is centred by its site mean:

$$\widetilde{\boldsymbol{x}}_i = \boldsymbol{x}_i - \boldsymbol{\mu}_{s(i)}. \tag{8}$$

This is a first-moment correction in the lineage of AdaBN [37] and the first-moment component of CORAL [30]; we evaluate its second-moment extensions (full CORAL, per-site ZCA; Supplementary Table S3).

**Counterfactual region ablation loss**

A counterfactual region ablation (CRA) loss acts as an attention-concentration regulariser on the graph branch, enforcing that removing the top-$m$ attended regions perturbs the prediction more than removing $m$ random regions, via a hinge on Kullback–Leibler (KL) divergences from the unperturbed output $p$:

$$\mathcal{L}_{\text{CRA}} = \text{ReLU}\big(D_{\text{KL}}(p \parallel p^{\text{rand}}) - D_{\text{KL}}(p \parallel p^{\text{abl}}) + \delta\big), \tag{9}$$

with $m$=3, $\delta$=0.1. We use "counterfactual" in the sense of region ablation, not as a formal causal claim (Discussion). A controlled ablation (Supplementary Note 7) shows it improves neither AUC nor attention concentration; we report it as a documented design choice, not a load-bearing component.

**Multi-scale patch fusion**

Two CHIME-MIL models $f^{256}$ and $f^{392}$ are trained identically; their per-slide probabilities are averaged with equal weight:

$$\hat{p} = 1/2\,(f^{256}(\mathcal{X}) + f^{392}(\mathcal{X})). \tag{10}$$

A leave-one-fold-out weight search gives an optimum $w^{*} \approx 0.47$ with a mean gain of only +0.0001 AUC over equal weighting, so flat 0.5/0.5 fusion is the practical late-fusion ceiling.

**Training and inference**

The total loss combines focal classification losses from all four heads with the CRA term:

$$\mathcal{L} = \mathcal{L}_{\text{focal}}(\hat{y}, Y) + \lambda_{\text{p}}\mathcal{L}_{\text{focal}}(\hat{y}_{\text{patch}}, Y) + \lambda_{\text{r}}\mathcal{L}_{\text{focal}}(\hat{y}_{\text{region}}, Y) \qquad +\lambda_{\text{g}}\mathcal{L} \tag{11}$$

with focal $\gamma$=2.0 [38] and $\lambda_{\text{p}}=\lambda_{\text{r}}$=0.2, $\lambda_{\text{g}}=\lambda_{\text{c}}$=0.3. Optimisation uses AdamW ($\eta$=5 × $10^{-5}$, weight decay $10^{-3}$) with early stopping on validation AUC (patience 10, maximum 40 epochs). At inference, the precomputed $\boldsymbol{\mu}_s$ is selected by the test slide's site identifier and subtracted; both resolution models forward through the pipeline and their probabilities are averaged (Equation (10)). The classification threshold is set at training time by maximising balanced accuracy on the hospital-stratified validation set and applied unchanged at test time.

**Evaluation protocol and statistics**

Generalisation is assessed by LOSO cross-validation: at each fold one hospital is held out for testing and the other four train the model, with no test-hospital data used in training or model selection. Early stopping and the decision threshold use a validation split drawn from the training hospitals. The headline configuration and the ablation rows marked SV use a hospital-stratified split (15% per hospital, label-stratified, capped at the median per-hospital count), so the largest site cannot dominate the best-epoch criterion; all other runs, including every baseline, use a label-stratified random 15% split of the pooled training slides. We report mean±s.d. AUC, accuracy and balanced accuracy across the five held-out folds (per-site reporting adds sensitivity, specificity, $F_1$ and positive predictive value), with malignant as the positive class.

Repeated runs of one configuration with the same seed do not give identical results unless the validation split and the CUDA settings are locked. The component ablation therefore uses a locked deterministic seed-42 reference run, in which the single-scale configuration with mean-centring and stratified validation reaches 0.8341. Two unlocked seed-42 runs of that configuration reach 0.8273 (the seed-42 member of the three-seed sweep) and 0.8289 (the default setting of the CRA sweep in Supplementary Note 5). This run-to-run spread (≈0.007) is as large as the individual mean-centring and stratified-validation gains, so single-run component differences are reported with confidence intervals and are not treated as established. Analyses of the single-scale model (onboarding, attention concentration, the CRA toggle and the branch decomposition) use the 0.8273 run.

AUC differences are tested with a paired slide-level bootstrap ($B$=10,000, seed 42, two-tailed, $\alpha$=0.05), resampling the same slides for both models. The bootstrap is used because AUC differences between correlated classifiers have no convenient closed-form distribution under site-heterogeneous sampling, and it makes no normality assumption. Two statistics are tested: the pre-specified pooled-LOSO AUC ($n$=5,036 slides) and, because the tables tabulate it, the five-fold-mean AUC (slides resampled within each held-out site; $n$=2,636, 1,183, 334, 685 and 198 for Sites A–E). For the headline the statistic is the mean over its three training seeds, so that the test concerns the tabulated number. The Benjamini–Hochberg correction [39] is applied within two pre-specified families against the headline: (i) six conventional learned-attention and transformer baselines the headline is expected to beat (ABMIL, CLAM-SB, CLAM-MB, TransMIL, DSMIL, RoFormer), and (ii) three foundation-model-ceiling probes that may match or exceed it (MaxPool at 256/392 px, ACMIL). Ablation steps are tested with the same paired bootstrap and reported with 95% percentile confidence intervals and unadjusted $P$ values; they are exploratory component attributions, not a pre-specified family. With $B$=10,000 the smallest resolvable two-sided $P$ is $2 \times 10^{-4}$; where no replicate crossed zero we report $P_{BH}$<0.001 (unadjusted: $P$<0.0002). The ± values are standard deviations across the five held-out sites, except the headline's ±0.0041, which is across its three training seeds, and the CAMELYON16 CHIME-MIL value, which is across its five training seeds.

**Baselines**

All baselines use the same LOSO folds, early-stopping rule (validation AUC, patience 10) and focal loss ($\gamma$=2.0) as CHIME-MIL, with the label-stratified validation split described above, on the UNI [17] (1,024-d) and/or GenBio-PathFM (4,608-d) feature tracks, both extracted at 20× under the same tiling. They are single-scale and, unless marked MC, unaligned, so comparisons with the headline compare recipes; recipe-matched comparisons are reported separately (Results). We compare against attention MIL (ABMIL [2]), CLAM-SB/MB [3], transformer MIL (TransMIL [4], RoFormer [5]), dual-stream DSMIL [8], pseudo-bag DTFD-MIL [9], and ACMIL [15] (the strongest attention-based baseline, evaluated with and without first-moment alignment to test transfer). A non-parametric global max pooling aggregator (element-wise maximum followed by a linear classifier, no learned attention) at 256 and 392 px serves as the foundation-model-ceiling reference.

**Implementation:** CHIME-MIL and all baselines are implemented in PyTorch 2.8.0 (UNI track on 4×RTX A5000 GPUs, GenBio headline on 4×H100 GPUs). Per-site means are a one-off CPU step. Hyperparameters are shared across variants and not tuned per fold (Supplementary Table S1).

**Use of generative AI:** No generative AI or large language model was used in the study design, the collection or annotation of the data, or the analysis or interpretation of the results; every GenBio-PathFM-track number reported in this paper is produced by the code listed under Data availability (the UNI-track baseline predictions were not retained). During manuscript preparation the authors used Anthropic Claude (a large language model) to draft and edit portions of the text and to assist with LaTeX formatting; the authors reviewed and edited the output and take full responsibility for the content.

**Data availability:** The whole-slide images analysed in this study are held privately by the authors' laboratory under the institutional review board approvals of the participating hospitals and are not publicly available because they contain protected clinical material; access requests may be addressed to the corresponding author and are subject to approval by the participating institutions and a data-use agreement. The derived patch features are available from the corresponding author on reasonable request. The CHIME-MIL model, training and evaluation code, the patch-feature loader, the site-mean alignment script and the late multi-scale fusion script are publicly available under the MIT licence at https://github.com/VanchhayNheng/CHIME-MIL. Pretrained CHIME-MIL checkpoints for all five leave-one-site-out folds, the remaining analysis code (the paired slide-level bootstrap tests, the baseline trainers, the higher-order alignment and attention-faithfulness evaluations and the patch-feature extraction script), the per-fold result files (probabilities, labels and thresholds), the precomputed per-site mean vectors and the exact split manifest for all 5,036 slides will be deposited in a Zenodo archive with a DOI issued at acceptance. Patch features are produced with the publicly released GenBio-PathFM model (genbio-ai/genbio-pathfm).

**Acknowledgements:** The authors thank the Departments of Pathology of the participating institutions for curating and releasing the archival whole-slide image cohorts used in this study.

**Author contributions:** N.V. conceived the study, developed the methodology with B.H.C. and H.-K.C., wrote the software, performed the formal analysis, prepared the visualisations and wrote the original draft. A.H., K.Z., M.A.I.M. R.I.S., and B.H.C. validated the results. W.-A.L., M.-S.C., H.-C.S., Y.P., N.-H.C., H.A., H.Ka. and H.Ki. provided the clinical cohorts and took part in the investigation. N.V., A.H., K.Z., M.-I.J., M.A.I.M., W.-A.L., M.-S.C., H.-C.S., Y.P., N.-H.C., H.A., H.Ka. and H.Ki. curated the data. H.-K.C. and H.C.K. supervised the work; H.C.K. administered the project and acquired the funding. All authors reviewed and edited the manuscript and approved the final version. (H.Ka., Haeyoun Kang; H.Ki., Hyunki Kim.)

**Funding :** This work was supported by the Institute of Information & Communications Technology Planning & Evaluation (IITP)-Innovative Human Resource Development for Local Intellectualization program grant funded by the Korea government (MSIT)(IITP-2024-RS-2024-00436773).

**Competing interests:** The authors declare no competing interests.

**Additional information:** Supplementary Information accompanies this paper (Supplementary Notes 1–12, Supplementary Tables S1–S15 and Supplementary Figure S1). Correspondence and requests for materials should be addressed to H.C.K.